\documentclass{aa}  

\usepackage{graphicx}
\usepackage{txfonts}
\usepackage{amsmath}
\usepackage{physics}
\usepackage{xcolor}
\usepackage{float}
\usepackage{hyperref}
\usepackage{threeparttablex}
\usepackage{siunitx}
\usepackage{orcidlink}
\usepackage{bm}
\begin{document}

   \title{Dynamic formation of multi-threaded prominences in arcade configurations}


   \author{V. Jer\v{c}i\'{c}
          \inst{1}\orcidlink{0000-0003-1862-7904}
          \,
          R. Keppens
          \inst{1}\orcidlink{0000-0003-3544-2733}
          }

   \institute{\inst{1} Centre for mathematical Plasma-Astrophysics, Celestijnenlaan 200B, 3001 Leuven, KU Leuven, Belgium}

   \date{Received date1 / Accepted date2}

 
  \abstract
   {High in the Sun’s atmosphere, prominences are plasma structures two orders of magnitude colder and denser than the surrounding corona. They often erupt, forming the core of violent and Earth-threatening coronal mass ejections. It is still unclear how these giant structures form and what causes their internal fine-structure and dynamics. How does mass and energy get exchanged with the lower layers of the Sun’s atmosphere?} 
   {We aim to understand the nature of prominences, governed by their formation process. How exactly does evaporation-condensation proceed, and how is the mass and energy exchange going on between the prominence and the regions where they are rooted, most notably the chromosphere and the transition region? We attempt to answer these questions.}
   {We use a state-of-the-art threaded prominence model within a dipped magnetic arcade. We solve the non-ideal magnetohydrodynamic (MHD) equations using the open source {\tt MPI-AMRVAC} MHD toolkit. Unlike many previous 1D models where magnetic field is assumed `infinitely strong', we study the full 2D dynamics in a fixed-shaped arcade. This allows for sideways field deformations and cross-field thermodynamic coupling. To achieve a realistic setup we consider field-aligned thermal conduction, radiative cooling and heating, wherein the latter combines a steady background and a localized stochastic component. The stochastic component simulates energy pulses localized in time and space at the footpoints of the magnetic arcade. We vary the height and the amplitude of the localized heating and observe how it influences the prominence, its threads, and its overall dynamics. }
   {We show with this work the importance of the random localized heating in the evolution of prominences and their threaded structure. Random heating strongly influences the morphology of the prominence threaded structure, the area, the mass the threads reach, their minimum temperature and their average density. More importantly, the strength of the localized heating plays a role in maintaining the balance between condensation and draining, affecting the general prominence stability. Stronger sources form condensations faster and result in larger and more massive prominences. We show how the condensation rates scale with the amplitude of the heating inputs and quantify how these rates match with values from observations. We detail how stochastic sources determine counterstreaming flows and oscillations of prominence threads.
   }
   {}

   \keywords{Sun: filaments, prominences --
                Sun: oscillations --
                methods: numerical
               }

   \maketitle
%

\section{Evaporation-condensation for prominences}
\label{sec:intro}
    Embedded within the million degree hot corona are solar prominences, structures of plasma two orders of magnitude colder and denser than the surrounding corona. Coronal rain is another form of condensed plasma with properties akin to prominences, albeit with clear morphological differences. Often, coronal rain is related to prominences through drainage. Prominence plasma can easily fall (drain) back to the chromosphere, in which case, the drained parts appear as coronal rain blobs. Numerous works have investigated how prominences or coronal rain blobs form and evolve. We know from observations that a prominence is a multi-structured, highly dynamic entity (on both small and large scale) \citep{Schmieder1991,Engvold1998,Lin2005,Hillier2012,Chen2014,Okamoto2016}. At scales of a few hundred kilometers or less, prominence threads lie at the limit of current instrument resolution. Consequently, much less is known about the lifecycles of these fundamental building blocks in prominences \citep{Lin2005}.
    
    For the formation of solar prominences, three scenarios are often considered \citep{Mackay2010}. In the \textit{injection model}, reconnection events at footpoints of coronal loops push (or inject) already "cold" plasma from the chromosphere directly into the corona. In \textit{formation by levitation}, cold and dense chromospheric plasma is levitated up to the corona, as a result of upward Lorentz forces \citep{Zhao2017,Zhao2019} following rearrangements in the magnetic field topology. Reconnection also plays a vital role here. Lastly, prominences can be formed via a process known as \textit{the evaporation-condensation process}. With localized heating present at the footpoints of a coronal loop or arcade, plasma evaporates into the corona, while local condensations form through a runaway process caused by increased radiative losses. On top of these three scenarios, several others have been demonstrated numerically: levitation-condensation in purely coronal volumes in both 2D and 3D \citep{Jenkins2021,Jenkins2022}, or even a plasmoid-fed prominence formation scenario that would naturally lead to filamentary structure aligned with the polarity inversion line \citep{Zhao2022}. In each of these multi-dimensional prominence formation models, the actual condensations result directly from thermal instability \citep{Parker1953, Field1965, Klimchuk2019, Antolin&Froment2022}, as quantifiable by linear magnetohydrodynamic spectroscopic analysis on the instantaneous loop profiles. Evidence for prominence formation via thermal instability is available in a number of observational studies \citep{Berger2012, Liu2012}. The necessary trigger and the resulting thermal instability have been studied with numerical simulations \citep{Antiochos1999, Muller2003, Muller2004, MendozaBriceno2005, Karpen2008, Antolin2010, Xia2016, Johnston2019, Claes2020, Zhou2020}. Many numerical studies looked into stochastic heating and the influence of its parameters; the heating scale, the pulse or interpulse duration, the background heating, the length of the loop, on the forming condensations \citep[cf.][]{MendozaBriceno2005, MendozaBriceno2006, Karpen2008, Johnston2019}. \cite{Mikic2013} pointed out the importance of nonuniformity along the loop (e.g. a change of the loops cross-sectional area or of the heating at the footpoints). \cite{Pelouze2022} extended that study by performing 9000 simulations exploring different heating parameters and different geometries. Whether coronal rain, a prominence or neither appears, depends on the combination of loop geometry and heating. 

     \begin{figure*}[ht]
      \centering
      \includegraphics[width=0.8\textwidth]{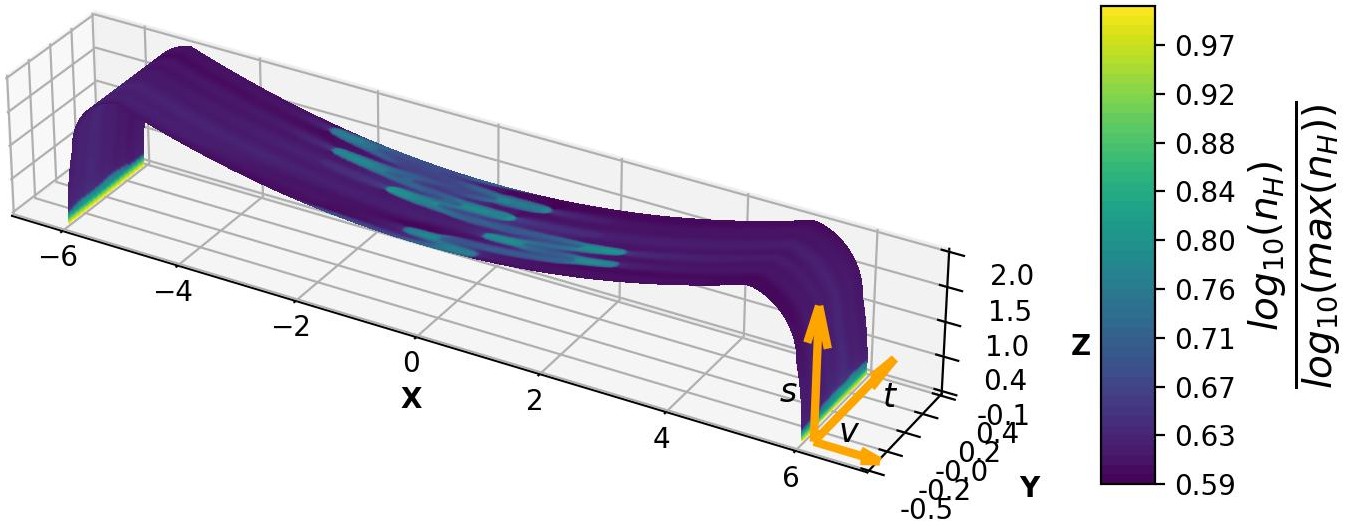}
      \caption{3D shape of our domain ($x$, $y$ and $z$ axis with units of $\times$10\,Mm) with over-plotted, normalized number density values at $t=261$\,min of the reference case. The coordinate system marked in orange describes $s$ and $t$ axis, along and transverse to the magnetic field lines, respectively. The $v$ axis is perpendicular to magnetic field lines (and not existing in our 2D domain). Note that the $x$, $y$ and $z$ axis are not to scale.}
        \label{fig:loop_shape}
    \end{figure*}
    
    One of the first 2D simulations to show successful formation of filament threads in a dipped magnetic arcade via thermal instability following stochastic footpoint heating was presented by \cite{Zhou2020}. The authors described the formation process of a prominence in a fixed arcade geometry, where they naturally obtained threaded fine structures. This clearly motivates further multi-dimensional modeling, and here we revisit this arcade setup and study the influence of the localized heating prescriptions. The localized heating adopted in the model must relate to the (not fully known) sources of solar coronal heating. \cite{Antolin2008} studied the difference in observational signatures between heating by Alfv{\'{e}}n waves and by nanoflares. In \cite{Antolin2010}, they showed that if the coronal loop is heated by Alfv{\'{e}}n waves, the coronal rain cannot form as the loop is heated uniformly. On the other hand, small localized heating events, representative of nanoflare reconnection heating, could successfully produce coronal rain. The works of \cite{Chae2003} and \cite{DePontieu2011} represent yet further evidence towards localized, intermittent heating drivers. The estimates of the mass supplied by the eruptive and jet-like events in the work by \cite{Chae2003}, found it to be sufficient to form a prominence, dominated by flows of 80-250\,km\,s$^{-1}$.
    
    The influence of stochastic heating, as supported by observations, remains poorly constrained. In a paper by \cite{Huang2021}, the authors showed with a 1D hydrodynamic model the importance of the height of the localized heating. This parameter can significantly influence the formation process, by differentiating between a prominence formed via evaporation-condensation or via an injection formation process. This motivates to experiment with that parameter in 2D scenarios. Exploring the different aspects of prominence formation in 2D and comparing the results with what we know from observations, is of paramount importance. In this manuscript, we will study the connection between impulsive heating and the threaded prominences found in the corona. We do this by varying the amplitude of the random heating pulses, and with it, changing the energy introduced into the system. Additionally, we change the height of the pulses with respect to the position of the transition region (TR). In Section~\ref{sec:methods} we describe the domain, the localized heating and numerical methods used. In Section~\ref{sec:results} we describe the results of our survey in detail and in Section~\ref{sec:discussion} discuss these results in more depth. Section~\ref{sec:conclusion} draws concluding remarks.


\section{Numerical Methods}
\label{sec:methods}

\subsection{Geometrical and physical configuration of the domain}

    The model we consider here is the evaporation-condensation model. We are interested in the behaviour of the prominence in an arcade configuration, viewing the arcade as a curved, fixed structure due to the strong magnetic field. Previous works mostly focused on arcade dynamics in the cross-sectional vertical plane, perpendicular to the Sun's surface \citep{Terradas2013, Keppens2014, Luna2016, Zhang2019, Liakh2021}. We model a solar prominence in a 2D dipped field arcade, and wish to include its fine, threaded-like structure. A stable magnetic configuration that can avoid threads draining away too fast is a magnetic arcade with a central dip (i.e. a concave upwards section). Hence, we use a similar magnetic field topology as in our previous study \citep{Jercic2022}. Figure~\ref{fig:loop_shape} describes how our domain is to be interpreted in a 3D setting ($x$, $y$ and $z$ axis). However, our numerical domain is a rectangular plane that straightens out the curved 2D arcade manifold, in a sense representing a top-down view over the structure we see on Fig.~\ref{fig:loop_shape}. The magnetic field is assumed strong, and bent in the given shape, where the dip would be a result of the gravitational force that opposes vertical tension. However, with the curved manifold of the arcade field of given shape, the dip does not deform vertically, and gravity only enters through its field-projected component. Information is allowed to propagate across field-lines ($s$ and $t$-direction on Fig.~\ref{fig:loop_shape}) and field line bending can happen in the manifold, allowing for fully 2D dynamics (both along and across field lines). Instead of using the $x$, $y$ and $z$ coordinates, we actually work with $s$ (along the field lines) and $t$ (transverse to the field lines) axis. In the rest of the paper we will continue referring to $s$ and $t$ axis as the $x$ and $y$ coordinates. In order to prevent condensations from easily draining over the shoulders of the arcade, we made the central section of the arcade slightly deeper than in our previous work. Therefore, the depth of the central section now measures 8\,Mm. The domain is in a hydrostatic equilibrium according to which we determine the pressure and density inside the domain (for details see Xia et al. 2011). The temperature distribution is determined according to the following equation,
    \begin{equation}
        T(z) =  T_{pho} + \frac{1}{2} (T_{cor}-T_{pho})\Bigg[1 + \tanh \bigg( \frac{z-h_{tra}}{w_{tra}} \bigg) \Bigg] \,,
        \label{eq:temperature}
    \end{equation}
    where $z$ corresponds to the $s$-direction on Fig.~\ref{fig:loop_shape}. $T_{pho}$ is the temperature of the photosphere at 6000\,K. $T_{cor}$ is the temperature of the corona at 1\,MK. $h_{tra} = 2.72$\,Mm and $w_{tra} = 250$\,km represent the height and width of the initial TR.

\subsection{Numerical strategy}
    
    We solve two-dimensional magnetohydrodynamic (MHD) equations (with no magnetic field or velocity component in the third, ignorable direction) with non-adiabatic effects, including: radiative losses ($n_Hn_e\Lambda(T)$), anisotropic thermal conduction ($\nabla \cdot (\kappa \cdot \nabla T)$), background and localized heating ($H$) as,
    \begin{eqnarray} 
    \label{eq:mass}
    \pdv{\rho}{t}+\nabla \cdot (\rho \textbf{v}) & = & 0  \,,\\ 
    \label{eq:momentum}
    \pdv{\rho \textbf{v}}{t} + \nabla \cdot \Bigg(\rho \textbf{vv} + p_{tot}\textbf{I}-\frac{\textbf{BB}}{\mu_0} \Bigg) & = & \rho \textbf{g} \,,\\
    \label{eq:energy}
    \pdv{e}{t} + \nabla \cdot \Bigg(e\textbf{v}-\frac{\textbf{BB}}{\mu_0}\cdot \textbf{v} + \textbf{v}p_{tot} \Bigg) &= & \rho \textbf{g}\cdot \textbf{v} + \nabla \cdot (\pmb{\kappa} \cdot \nabla T)  \, \nonumber \\
    & & - n_Hn_e\Lambda(T) + H \,, \\
    \label{eq:induction}
    \pdv{\textbf{B}}{t} + \nabla \cdot (\textbf{vB} - \textbf{Bv}) & = &0\,,
    \end{eqnarray}
    \noindent where $p_{tot}=p+\frac{B^2}{2\mu_0}$ is the total pressure, corresponding to the sum of thermal pressure $p$ (calculated via the ideal gas law $p=2.3n_Hk_BT$) and the magnetic pressure. The magnetic field is initially uniform ($\sim$\,10\,G) and oriented along the $x$ axis in the domain. The value of gravity at the solar surface is 274\,m\,s$^{-2}$. The geometry of the magnetic arcade determines the distribution of the field-aligned gravity component, i.e. $\textbf{g}$ is gravity whose $x$-component corresponds to a fixed vertical gravity component that is locally projected along the prescribed magnetic arcade shape. Also, $\rho$, $\textbf{v}$, $e$, and $\textbf{B}$ are plasma density, velocity, total energy density and the magnetic field, respectively. $\kappa$ is the coefficient of thermal conductivity parallel to the magnetic field, as usual taken to be equal to 8$\times$10$^{-7}$\,T$^{5/2}$\,erg\,cm$^{-1}$\,s$^{-1}$\,K$^{-1}$ \citep[Spitzer conductivity coefficient,][]{Spitzer2006}. The perpendicular conductivity is considered negligible and hence we do not take it into account. The plasma we simulate is fully ionized, and the hydrogen to helium abundance ratio is 10:1. Following that, the plasma density $\rho$ corresponds to 1.4$m_pn_H$, where $n_H$ is the number density of the plasma and $m_p$ is the proton mass. 
    
    Equations \ref{eq:mass}-\ref{eq:induction} are solved using an open-source MHD simulation code, the MPI\ Adaptive Mesh Refinement Versatile Advection Code ({\tt MPI-AMRVAC}\footnote{http://amrvac.org/}) \citep{Keppens2012, Porth2014, Xia2018, Keppens2021}. For the spatial discretisation, we employed a Harten-Lax-van Leer (HLL) approximate Riemann solver \citep{Harten1983} combined with a third order, asymmetric shock-capturing slope limiter as proposed in \cite{Cada2009}. To ensure stability, the Courant number we used is 0.8. We can rely on this Courant number, since the otherwise restrictive conduction timestep condition is overcome by a super-timestepping RKL2 scheme proposed by \cite{Meyer2012} \citep[for further discussion see][]{Xia2018}. Time discretisation is done with a five-step (strong stability preserving) fourth-order Runge-Kutta method \citep{Spiteri2002}. To maintain the divergence of the magnetic field as approximately zero we used the upwind constrained transport method by \cite{Gardiner2005}, available as one of the many divergence control measures in {\tt MPI-AMRVAC}. Radiative losses scale with $\rho^2$ and are determined by $\Lambda(T)$, the function of temperature representing radiative loss per unit mass. $\Lambda(T)$ is prescribed with a cooling curve and {\tt MPI-AMRVAC} has multiple choices. In this work we used the one described by \cite{Schure2009}, where the temperature treatment below 10000\,K was added according to \cite{Dalgarno&McCray1972}. Additionally, we enforced a temperature minimum throughout the grid, at a value of 4170\,K, to avoid negative pressure issues in radiative cooling instabilities. On the full domain of 150\,Mm~$\times$~8\,Mm we employed adaptive mesh refinement (AMR), but only used 2 levels of AMR (including the base level with 520$\times$100 cells), resulting in a resolution of 144$\times$40\,km at the highest level of refinement. Refinement is set based on the Lohner prescription for strong gradients in the density field. Following the bootstrap measure described in \cite{Hermans2021}, we can avoid encountering too-small pressures or densities (and hence timesteps) by using the "replace" method, controlled in  the parameter file through $small\_pressure=10^{-5}$ and $small\_density=10^{-3}$ (expressed in their dimensionless form). 
    
    The boundary conditions along the $y$ direction (across the arcade) are periodic, for the sake of stability and simplicity. We include the transition region and the chromosphere in the simulation. In order to have a physically correct energy transfer between the two systems, we employ the "transition region adaptive conduction" (TRAC) method \citep{Johnston2020, Zhou2021} and we describe it in more detail in Sec~\ref{sec:TRAC_resolution}. Along the $x$ direction, the magnetic field is extrapolated at the footpoints, while the velocity is reflective, mimicking the effects of the dense photospheric regions. Pressure and density are initially calculated according to the hydrostatic equilibrium, with the footpoint values fixed for all timesteps \footnote{This is the same prescription as in our previous paper, \cite{Jercic2022}, but a small change in the interpolation strategy leads to different values of coronal temperature and pressure.}.
    
\subsection{Heating prescription}
\label{sec:heating_prescription}
    
    The heating in this work is composed of two components, the time independent background heating that is needed to balance initial radiative losses and maintain the hot corona, and the localized heating that varies in space and time. For the background heating we use a simple exponential function dependent on the $x$ coordinate and valid for every $y$,
    \begin{equation}
        H_0 = \begin{cases}
        E_0\exp(-x/H_m), & x < L_x/2 \\
        E_0\exp[-(L_x-x)/H_m], & L_x/2 \leq x < L_x
        \end{cases}
    \end{equation}
    where $L_x$ marks the full length of the domain in the $x$ direction (150\,Mm), $E_0=5\times10^{-5}$\,erg\,cm$^{-3}$\,s$^{-1}$ is the amplitude and $H_m=L_x/2$ is the heating scale length in the $x$ direction. The same type of background heating was already used in multiple other 1D and multi-D studies \citep[cf.][]{Xia2011,Zhang2013, Xia2016,Zhou2017, Zhou2020}. To form our localized heating we simulate a series of events that are random and localized in time and space. The idea is to simulate the so-called "nanoflare storm" \citep{Lionello2013}, i.e. a series of small reconnection events at the footpoints of coronal loops where the magnetic field lines supposedly twist and braid \citep{Parker1988}. In order to simulate this we created a localized heating similar to the one used in \cite{Antolin2010}, 
    \begin{align}
    \label{eq:Hi}
        & H_i(t,x,y) =\nonumber \\
        &\begin{cases}
        E_1\sin\bigg(\frac{\pi(t-t_i)}{\delta t_i}\bigg)\exp\bigg(\frac{-\abs{x-x_i}}{x_h}\bigg)\exp\bigg(\frac{-\abs{y-y_i}}{y_h}\bigg), \hspace{0.3cm} t_i < t < t_i + \tau_i \\
        0, \hfill \text{otherwise}
        \end{cases}
    \end{align}
    where $\delta t_i$ is the pulse duration. The amplitude of this heating is defined as: $E_1=A(1+\frac{\tau_i}{\delta t_i})(1+\frac{y_h}{L_y})$ where $\tau_i$ is the interpulse duration time and $L_y$ is the full extent of the domain in the $y$ direction (in total 8\,Mm). The amplitude $A$ is increased by two factors in the brackets in order for the heating, localized in time and space, to be comparable to a successful 1D simulation with steady heating. If the localized pulses are too weak, the resulting perturbations in the corona are not enough to trigger thermal instability. A similar approach for maintaining an energy release of a steady 1D heating equivalent to the one localized in time, was used by \cite{Johnston2019}. For a reference case we use $A=1.8\times10^{-2}$\,erg\,cm$^{-3}$\,s$^{-1}$. Eventually, the total heating due to nanoflares is then represented by,
    \begin{equation}
    \label{eq:Htot}
        H=\sum_{i=1}^{n} H_i(t,s)\,.
    \end{equation}
    The pulse duration time, $\delta t_i$ is 300\,s$\pm$75s \citep{Zhou2020}. The interpulse duration, $\tau_i$ is a set of random numbers created in the range between 100 and 300\,s with Fortran 90 intrinsic function \texttt{random\_number}. The fact that our interpulse and pulse duration are short ($<$\,400\,s) is analogous to having a steady heating \citep[cf.][]{Johnston2019}. Choosing the pulse duration is not as trivial as it may seem and values from quite short \citep[e.g. 20\,s in][]{Karpen2008} to quite long \citep[e.g. 8000\,s in][]{Johnston2019} have been used in different studies. Parameters $x_h$ and $y_h$ are heating scale lengths in the $x$ and $y$ directions, respectively. Both are set to 1.5\,Mm. The positions $(x_i,y_i)$ of the heating pulses is random, also created with Fortran 90 intrinsic function \texttt{random\_number}. Along the $x$-axis, the pulses can appear around the TR in the limited range of 2\,Mm (see also Sec.~\ref{sec:Loc_heating_influence}). While in the $y$ direction, it is located in the range between $y_1=-2.5$\,Mm to $y_2=2.5$\,Mm. The same procedure of creating and setting up pulses was done along both footpoints of the magnetic arcade. Different random events are created for each footpoint and hence footpoints are treated independently, resulting in asymmetric heating. The cases of different amplitudes studied here, all have the same random seeds, only the amplitude has been modified. As for the different heights, new random seeds were created for each case (\textit{high} and \textit{low}), but all still with the same value of amplitude $A$. 
    
    Considering that the domain represents an arcade composed of different geometrical parts and that on top of that we impose a steady background heating and have thermal conduction, initially the system is not in complete numerical nor force equilibrium. Therefore, we first let the system relax, imposing only the background heating. We let the relaxation phase last for about 50\,min of physical time, after which the TR height is close to 2\,Mm while the maximum velocity along the $x$-axis in the domain (the coronal part) is still close to 13\,km s$^{-1}$ (it reduces even more if we use higher resolution grids). Due to the type of computational domain we use (periodic boundary conditions with high density gradients at the footpoints) any velocity perturbation that starts in this domain can never really leave. As we will anyway induce perturbations in the domain which are of the same or higher order of magnitude there is no need to fully dampen these velocities. After the relaxation phase, we reset time to $t=0$. At that moment, we include the localized heating, so the first $t_i$ on the left footpoint is set at $t=0$ while on the right footpoint, the first pulse happens 5.5\,min later.

\subsection{Influence of the resolution}
\label{sec:TRAC_resolution}
    Given that we include the chromosphere, high enough resolution is needed to avoid erroneous densities \citep{Bradshaw&Cargill2013}. Furthermore, considering every thread has a transition region bordering it, one needs to properly resolve the Field length. Field length was described by \cite{Field1965} and later dubbed as Field length by \cite{Begelman&McKee1990}. It characterizes a length scale given as a square root of the ratio of thermal conduction and optically thin radiative losses at temperature $T$. The detailed balance of radiation and conduction can influence the width of the formed condensations. Field length is defined as
    \begin{equation}
        \lambda_F = \Bigg( \frac{\kappa(T)T}{n^2\Lambda(T)} \Bigg)^{1/2} \,.
    \end{equation}
    Its role is discussed in simulations of galaxy clusters in \cite{Sharma2010} or for prominences in works by \citep{KoyamaInutsuka2004, Kaneko2017}. \cite{Hermans2021} looked at thermal instability evolutions in 2D setups, without thermal conduction included, but they pointed out how the Field length must inevitably be locally underresolved in actual multidimensional condensation transition regions. In a pure hydrodynamic case of isotropic thermal conductivity, the Field length needs to be resolved by a few grid cells in order for the simulation to converge \citep{KoyamaInutsuka2004}. In the coronal case of anisotropic conductivity, even the Field length perpendicular to the magnetic field may need to be resolved. As it is not yet computationally feasible to resolve such small lengths, we used the "transition region adaptive conduction" (TRAC) method \citep{Johnston&Bradshaw2019,Johnston2020}, recently implemented in {\tt MPI-AMRVAC} for true multi-dimensional applications \citep{Zhou2021}. From multiple approaches of this method implemented in {\tt MPI-AMRVAC}, we used the \textit{mhd\_trac\_type=2}. Type 2 is the localized TRAC (LTRAC) first proposed by \cite{Iijima2021}. The TR is broadened using the information only of the nearby grid points. TRAC numerically smoothens the area where the Field length decreases drastically (most notably the TR), but ensures proper capturing of mass evaporation and energy exchange between the relatively cold TR and hot corona. Besides the TR between chromosphere and corona, we also have the prominence-corona transition region (PCTR) at each thread edge. The resolution used in this study is relatively low, 144\,km $\times$ 40\,km on the highest refinement level. As a result, the width of the threads that form is influenced by resolution (higher resolution results in thinner threads). 

\section{Results}
\label{sec:results}
    We present here the results of 6 different simulations where we follow the evolution in our domain for a total of 5\,hr. The models were not run long enough to observe the total prominence evolution and we address this point later on in greater detail. The condensations take $\approx$\,2\,hr to form. For the remaining 3\,hr, we study the evolution of the prominence threads for different parameters of the localized heating. We start with a reference case, followed by cases wherein we varied the parameters of the localized heating in Eq.~(\ref{eq:Hi}). The following analysis was performed in a boxed region of our domain encompassing the $x$ length from 20 to 130\,Mm and the full extent in the $y$ direction. 
    In order to extract the threads we use a number density threshold: $n_H > 7\times10^9$\,cm$^{-3}$. Everything within the chosen box but below the threshold density for prominences is considered coronal. The average number density of that coronal area at $t=0$\,min (just after the relaxation phase) is 2.97$\times$10$^8$\,cm$^{-3}$.
    
\subsection{Formation – the reference case}
\label{sec:reference_case}
    \begin{figure}
      \centering
      \includegraphics[width=\hsize]{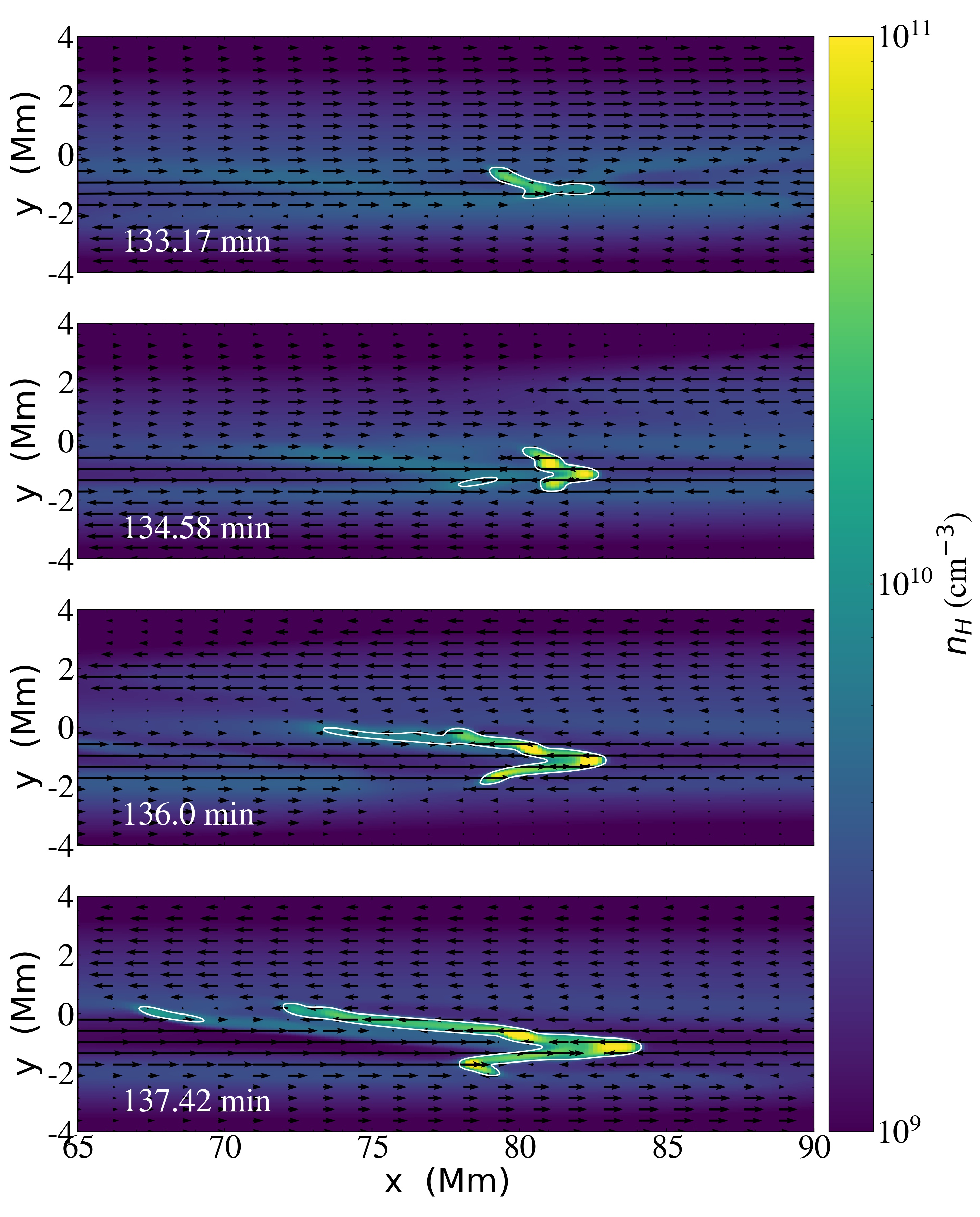}
      \caption{Number density of the reference case in the first 4\,min of the condensation (first thread) forming, with the arrows representing the $x$ component of the velocity and the white contours marking the thread.}
        \label{fig:formation}
    \end{figure}
\begin{figure*}
      \centering
      \includegraphics[width=\textwidth]{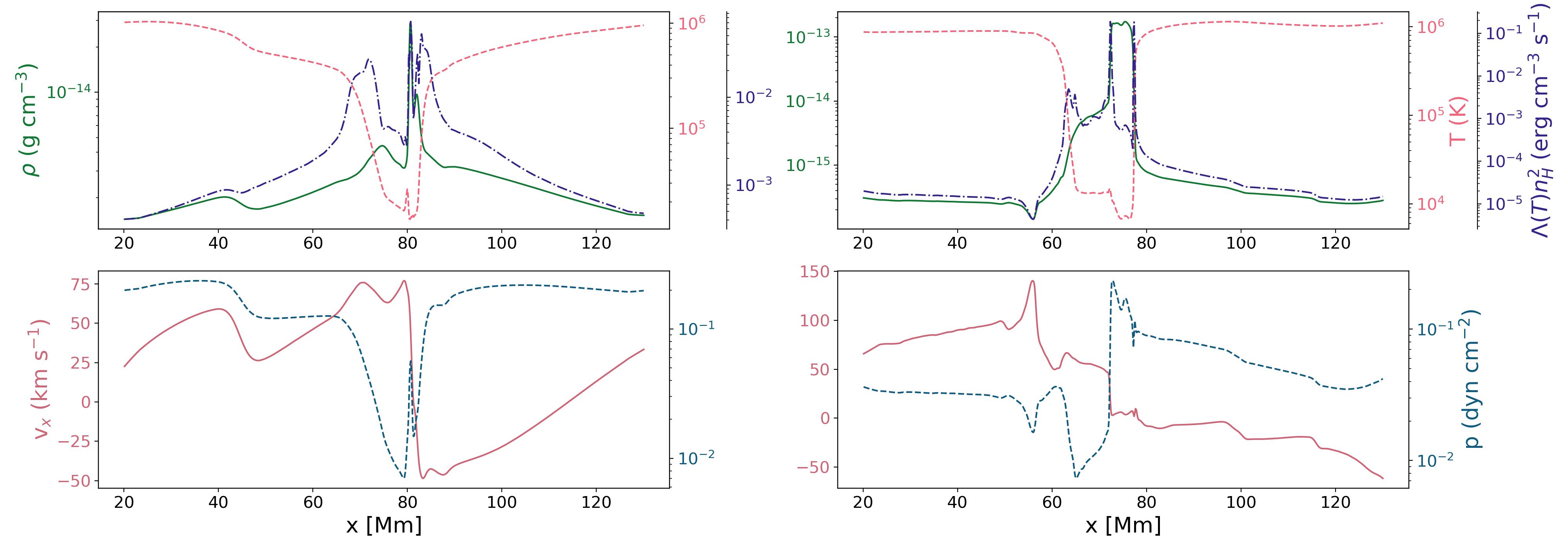}
      \caption{$x$-variations of quantities at $t=133$\,min (left panels, when the first condensation forms) and $t=204$\,min (right panels). The cut along $x$ at $t=133$\,min is at $y=-1.2$\,Mm and the cut along $x$ at $t=204$\,min is at $y=-0.8$\,Mm. On the top panels the full green line is density, the dashed red line is temperature and the dashed-dotted indigo line is the radiative losses. On the bottom panels the full red line is the $x$ component of the velocity and the dashed blue line is thermal pressure.}
        \label{fig:xlineplot}
    \end{figure*}
\begin{figure*}
      \centering
      \includegraphics[width=\textwidth]{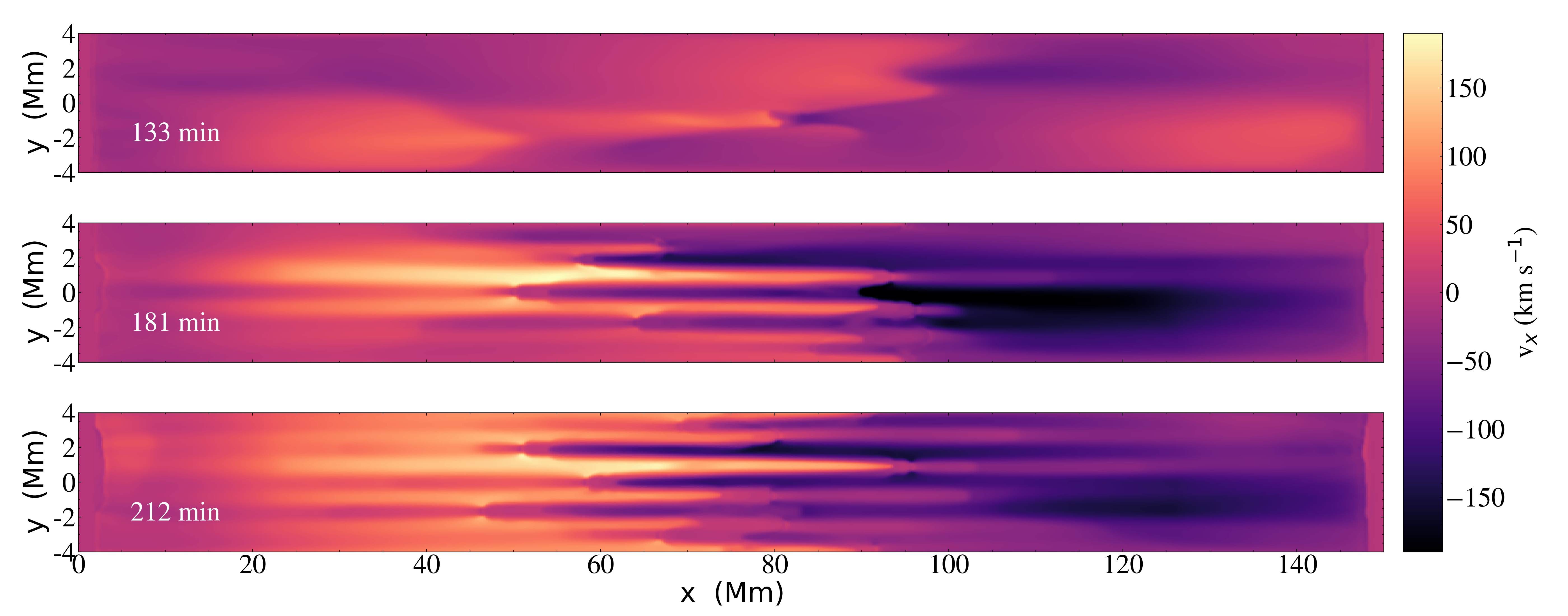}
      \caption{$x$ component of the velocity in the full domain of the reference case at three different moments during the evolution. The top panel is at the same time first shown in zoomed view in Fig.~\ref{fig:formation}.}
        \label{fig:1Avelocities}
    \end{figure*}

    Figure~\ref{fig:formation} shows four snapshots of the number density plot with velocity quivers indicating the flow around the region where the condensation first appears (with the contours defined by the aforementioned $n_H$ threshold). From the plots we see that the thread forms at the position where oppositely directed flows collide. At this point in time and space there is enough matter for the radiative cooling to overcome the otherwise stabilizing thermal conduction in a thermal instability. The temperature and gas pressure are perturbed such that they both decrease significantly. Because the gas pressure drops, more matter is pulled in. As more matter is pulled in, the density starts increasing. Consequently, the radiative losses (dependent on $n_H^2$) also increase, which further decreases the temperature. Thermal instability sets in and initiates runaway cooling. Once the condensation starts forming, it grows in length and also extends along the $y$ direction. In the first 10\,min the area the threads occupy grows to 1.42$\times$10$^{17}$\,cm$^2$. The threads that eventually form are all connected. In general, they follow the magnetic field lines with a small but significant deviation (up to approximately 2$^{\circ}$). The densest parts of the threads (the edges) follow the field lines exactly. Moving away along the threads from its edge, the density decreases slightly, and the thread is then inclined away from the magnetic field line just enough to be connected to the neighbouring thread. In the upper panels of Fig.~\ref{fig:xlineplot} we plot density $\rho$, temperature $T$ and radiative losses $n_H^2\Lambda(T)$ at $t=133$\,min and $t=204$\,min, while in the bottom panels we plot the $v_x$ component of velocity and thermal pressure $p$, again for the same time moments. Each panel shows  their variations along $x$ (along the arcade) at $t=133$\,min, taken at fixed  $y=-1.2$\,Mm along which the first thread appears (Fig.~\ref{fig:formation}). The cut along $x$ at $t=204$\,min is for fixed $y=-0.8$\,Mm, exactly through another thread we find at that moment. From the two panels on the left we can follow the condensation process. We notice how at the position where the flows collide the density peaks, which points to the importance of these flows in the formation and evolution of condensations. At the same time, radiative losses also peak there, hence the temperature experiences a significant drop. Looking at the right panels of Fig.~\ref{fig:xlineplot}, representing a later stage in the evolution of the thread, we notice more complex changes along it. The threads are the densest at the edges, where we see the density in our top right plot reaching values approximately 10$^{-13}$\,g\,cm$^{-3}$. At the edges of that dense region the radiative losses show a peak, representing the thread's transition region. On the left of that main peak, the region is still relatively dense enough to be considered part of the thread. Because the thread is slightly inclined to the magnetic field (which is mainly along $x$), with this plot we do not capture exactly its other outer edge. Moving our attention back to the pressure and velocity in both the bottom panels, it is interesting to observe how the thermal pressure at $t=133$\,min (left panel) is approximately identical at left and right of the forming condensation. While, later on, at $t=204$\,min (right panel) the pressure shows differences at the left and right side of the thread, creating a pressure gradient obviously contributing to the motion of the thread.
    In Fig.~\ref{fig:1Avelocities} we show the $v_x$ velocity components of the reference case at three different moments during the evolution of the threads. At 133\,min we have the first condensation appearing (topmost panel of Fig.~\ref{fig:formation}). We see flows coming from both sides of the domain in bands of different widths. They collide and flow past each other creating a pattern of counter-streams.  At 181\,min the velocity along the $x$ direction inside the region of 20 to 130\,Mm is in the range of -210.84\,km\,s$^{-1}$ and 190.20\,km\,s$^{-1}$. At this point and the next one shown at $t=212$\,min, we clearly see a banded flow where the width of the bands depends on the width of the threads that have already been formed. Unidirectional flows coming from one of the two sides of the domain collide with these threads, get transmitted through them and become mixed, while there are also longitudinal oscillations the threads themselves exhibit.

\subsection{Influence of the localized heating}
\label{sec:Loc_heating_influence}

    In order to understand how localized, random in time and space heating can influence the threads that condense, we varied the pulses amplitude $E_1$ in Eq.~(\ref{eq:Hi}) and the footpoint regions where the pulses happen. The four cases of different amplitude are: the 1A case (i.e. the reference case described in Sec.~\ref{sec:reference_case}), 0.75A, 1.5A and 2A. By changing the amplitude of the source we in fact change the energy of each pulse. As for changing the height of the pulses, we have three different cases, \textit{low}, \textit{middle} (same as 1A) and \textit{high}. The pulses in the \textit{low} case happen between 0.5 and 2.5\,Mm in the left footpoint and 147.5 and 149.5\,Mm in the right one. The region for the \textit{middle} case is 1\,Mm higher and for the \textit{high} case another 1\,Mm more. Considering the TR region has a height of around 2\,Mm after the relaxation phase, the pulses in the \textit{low} case happen mostly below it. The pulses in \textit{middle} case are around the TR and in the \textit{high} case, above it. However, due to the localized heating the TR will move around, depending on when and where the pulses happen. As a result, the described position of the pulses with regards to the TR is only an average measure.

\subsubsection{Mass and Area}
\label{sec:mass_area}
    
\begin{figure}
      \centering
      \includegraphics[width=\hsize]{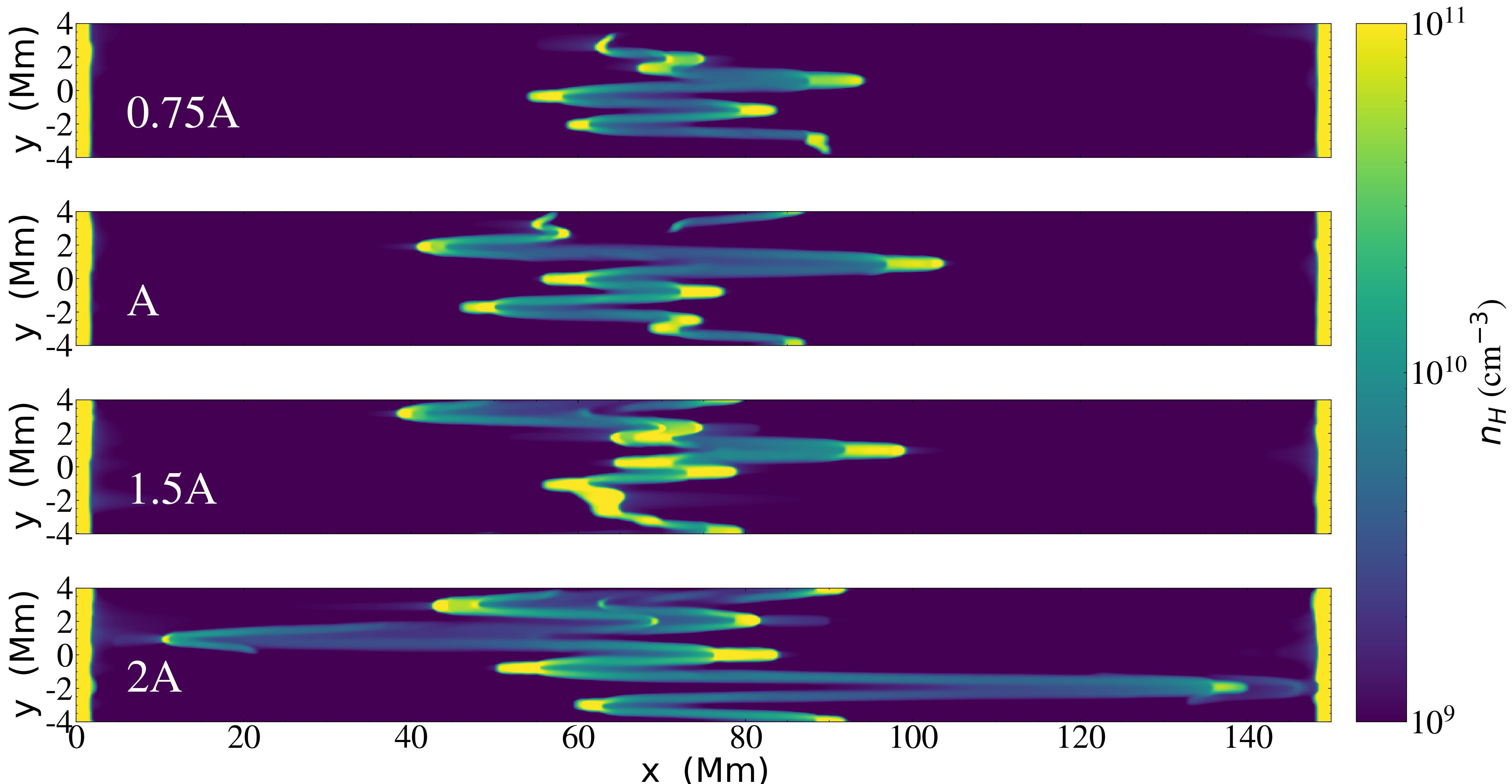}
      \caption{Morphology differences in the four cases with different pulses (at $t=204$\,min).}
        \label{fig:morphology_energies}
    \end{figure}

\begin{figure}
      \centering
      \includegraphics[width=\hsize]{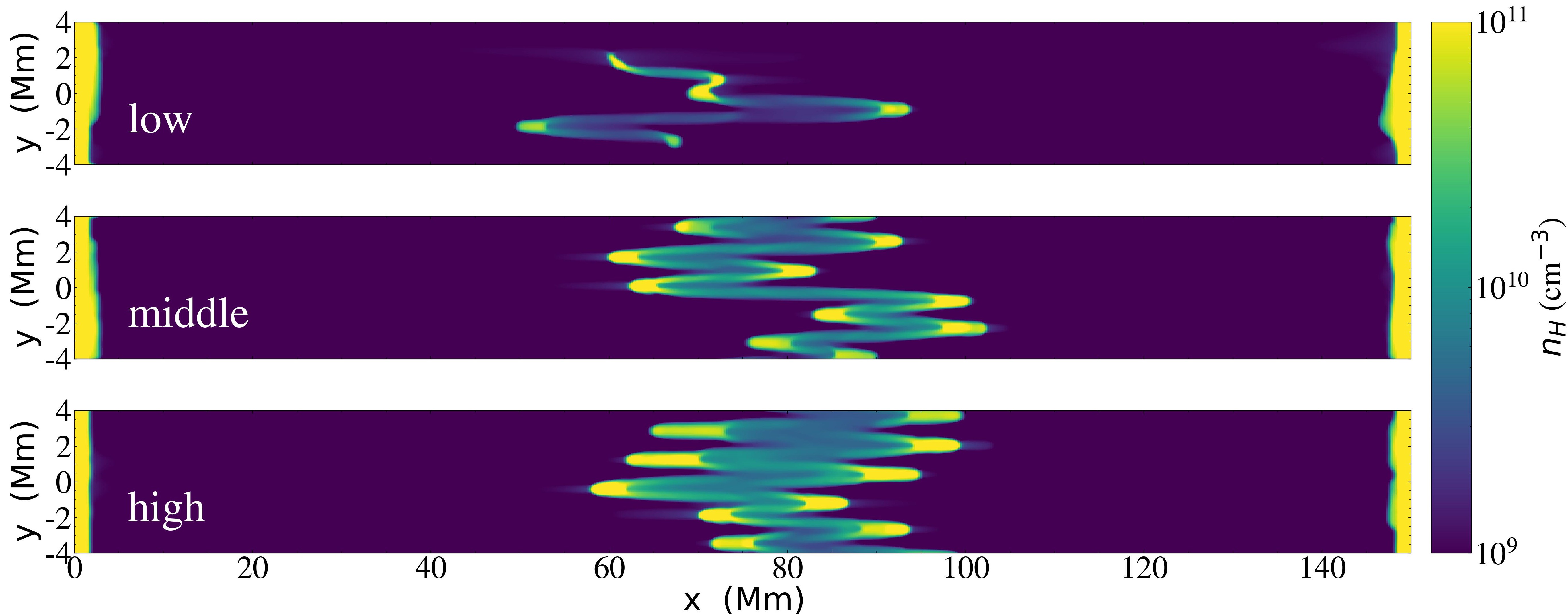}
      \caption{Morphology differences in the three cases with different heights of pulses (at $t=245$\,min).}
        \label{fig:morphology_heights}
    \end{figure}

    The flows (i.e. the counter-streams) have a strong influence on where the condensation will form and how it will behave. These differences lead to further variations resulting in noticeable differences in the condensations that eventually form. In Fig.~\ref{fig:morphology_energies} and~\ref{fig:morphology_heights} we show these differences in the morphology of the threads for each case. For example, for the 0.75A case the threads are shorter (10-30\,Mm) in comparison to the 2A case, where the threads extend from the center all the way to the edges of the domain. The threads become stretched and elongated as a larger amplitude case causes more extensive displacements, and the plasma can reach the shoulders of the arcade, where it splits and drains out of the concave upward section, which can be interpreted as manifestation of coronal rain, or in other words mass drainage \citep{Bi2014,Jenkins2018,Jenkins2019,Fan2020,Xue2021}. Other differences can be seen comparing the \textit{low} and \textit{middle}, or the \textit{low} and the \textit{high} cases. The \textit{low} case has noticeably less threads and doesn't extend yet along the full extent in the $y$ direction.
    
    \begin{figure}
      \centering
      \includegraphics[width=\hsize]{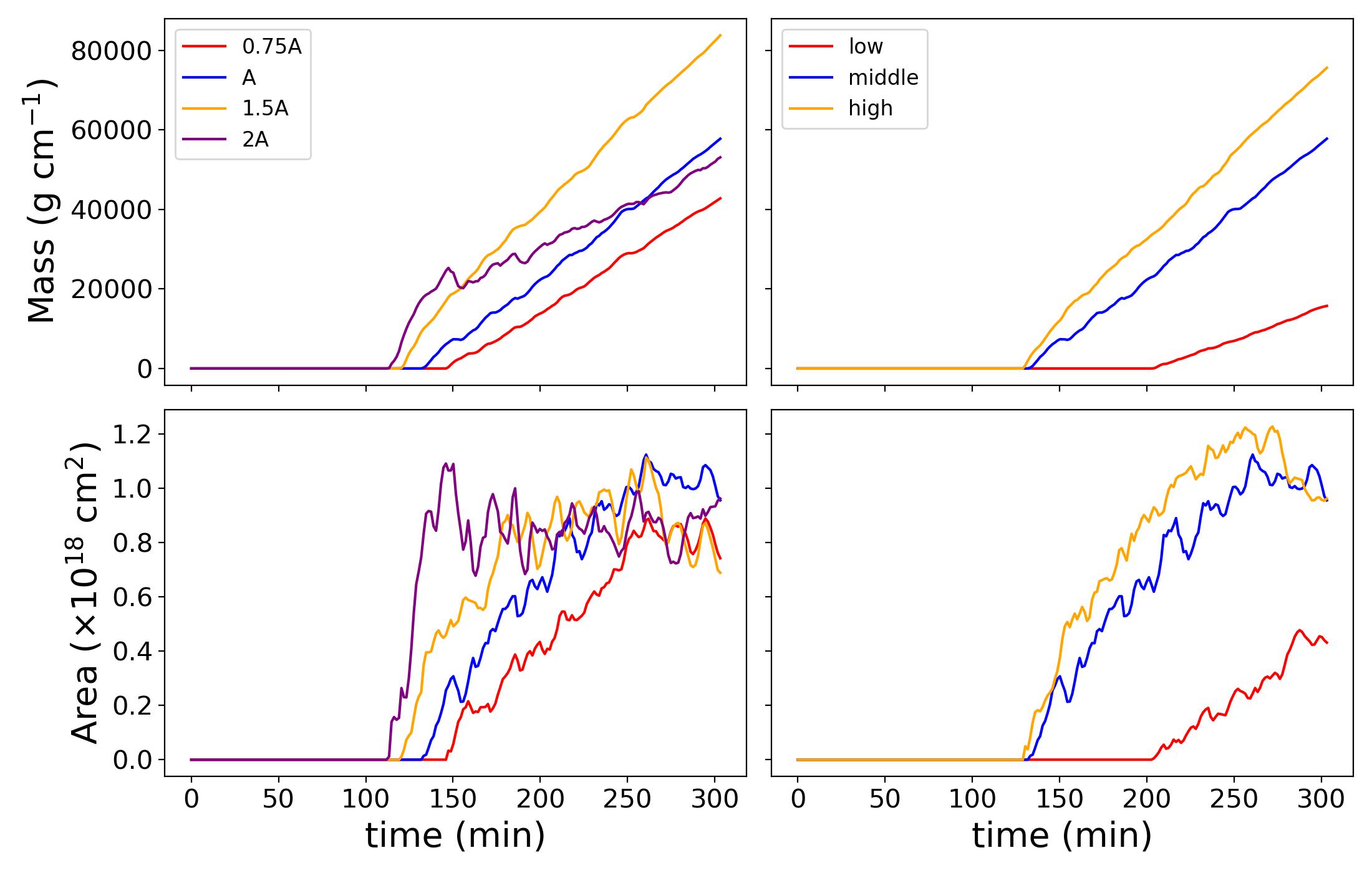}
      \caption{Change in time of the total area (bottom) and mass (top) for all the cases studied here. Left panels vary the amplitude (as in Fig.~\ref{fig:morphology_energies}), right panels the pulse height (as in Fig.~\ref{fig:morphology_heights}).}
        \label{fig:total_A_m}
    \end{figure}
\begin{figure}
      \centering
      \includegraphics[width=0.9\hsize]{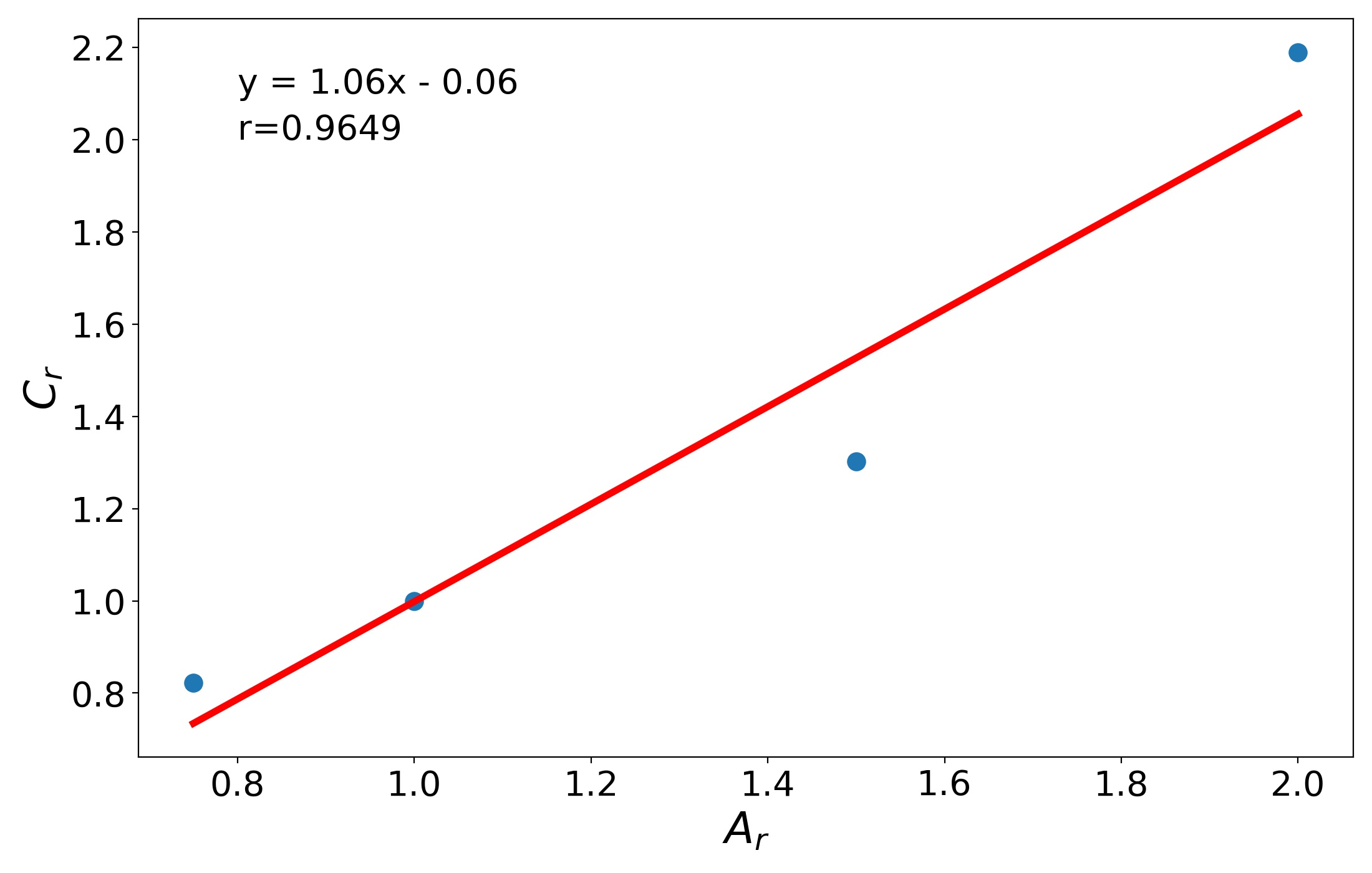}
      \caption{Linear relation of the normalized condensation rates, $C_r$ (Table~\ref{table:cond_rate}) with the perturbation amplitudes, $A_r$ (scaled to the reference case).}
        \label{fig:lin_condrates_amplitudes}
    \end{figure}

    Figure~\ref{fig:total_A_m} shows the change of the threads mass and area for the varying amplitude and pulse locations. From the top panels it is clear that the mass growth continues linearly from the moment of the first condensation. The energy pulses, that drive the evaporation and consequently drive the buildup of plasma in the corona, are continuously present during our simulated evolution and regulate the mass accumulation. We calculated the condensation rates by fitting this growth of mass in time with a linear function. The resulting slope coefficients are given in Table~\ref{table:cond_rate}. Further on, the normalized condensations, $C_r$ scale linearly with the ratio of amplitudes to the reference case ($A/A_{ref}=A_r=$\,0.75, 1, 1.5 and 2, see Fig.~\ref{fig:lin_condrates_amplitudes}). A linear fit provides a relation of $C_r=1.06A_r-0.06$ with a correlation coefficient of 0.96 for condensation rate $C_r$ as function of amplitude ratio $A_r$. As expected, the higher the amplitude of the pulse, or the more energetic the pulse, the more plasma is able to evaporate into the corona. As a result, we see more massive prominences. 
    As mentioned, case 2A is specific as it experiences drainage. That fact makes its condensation rate lower than for 1.5A case. Taking into consideration only the first 36.83\,min when the total mass of 2A case exhibits a local peak, we find a condensation rate of 12.24\,g\,cm$^{-1}$\,s$^{-1}$. In Table~\ref{table:cond_rate} we can also inspect the condensation rates for pulses that are at different heights (low-middle-high). Clearly, the higher the pulse is in the atmosphere, the higher the condensation rate is and hence a more massive condensation is formed in the same amount of time. 
    
    Another important parameter is how the area of the threads changes in time. From the bottom panels of Fig.~\ref{fig:total_A_m} we show that in all four cases of differing amplitudes the area increases sharply in the very beginning. That sharp increase is of greater magnitude and duration for a pulse of higher amplitude. Similar behaviour is found with the three cases of different pulse heights. In general, the higher the source region is located within the atmosphere the more dramatic area growth is seen, particularly in the initial phase of the formation of condensations. There is less difference between the \textit{middle} and \textit{high} cases than between either and the \textit{low} one. After the initial sharp increase, it seems the area growth stabilizes and the area increases with a steady rate. The prominence in 0.75A case experiences an increase in its total area by 4.77$\times$10$^{17}$\,cm$^2$ during the 80\,min between $t=160.08$\,min and $t=240.83$\,min. In the same period the 1A and 1.5A cases increase their area by 5.99$\times$10$^{17}$\,cm$^2$ and 3.72$\times$10$^{17}$\,cm$^2$, respectively. The 2A case is specific as it is the only one that experiences drainage, appearing clearly visible as a considerable drop in the area value around 150\,min. A similar steady increase in area is also seen for the cases of sources with different heights. The \textit{high} case experiences an increase in area by 5.97$\times$10$^{17}$\,cm$^2$ during the 80\,min interval (from $t=160.08$\,min to $t=240.83$\,min). The \textit{low} case which shows the slowest and the smallest increase, changes by 3.46$\times$10$^{17}$\,cm$^2$ during the 70\,min period between $t=211.08$\,min and $t=281.92$\,min. From Fig.~\ref{fig:total_A_m} we find that in each case, once the threads reach an area close to or above 1$\times$10$^{18}$\,cm$^{2}$, they stop growing. Around this value the threads (with the currently used resolution) start merging. This saturation in area value is not dependent on the energies introduced into the system nor on the positions of the source. This merging of the threads, seen as cessation of area increase, can be explained by the specific magnetic field setup shown in Fig.~\ref{fig:loop_shape}: the deep concave upwards section adopted here causes a relatively strong gravity component, causing compression of the threads. After they accumulate enough mass, threads then start merging across field lines (at times beyond those shown in all 2D figures so far).

\subsubsection{Density and temperature}
\label{sec:numdensity_temp}
    
\begin{figure}
      \centering
      \includegraphics[width=\hsize]{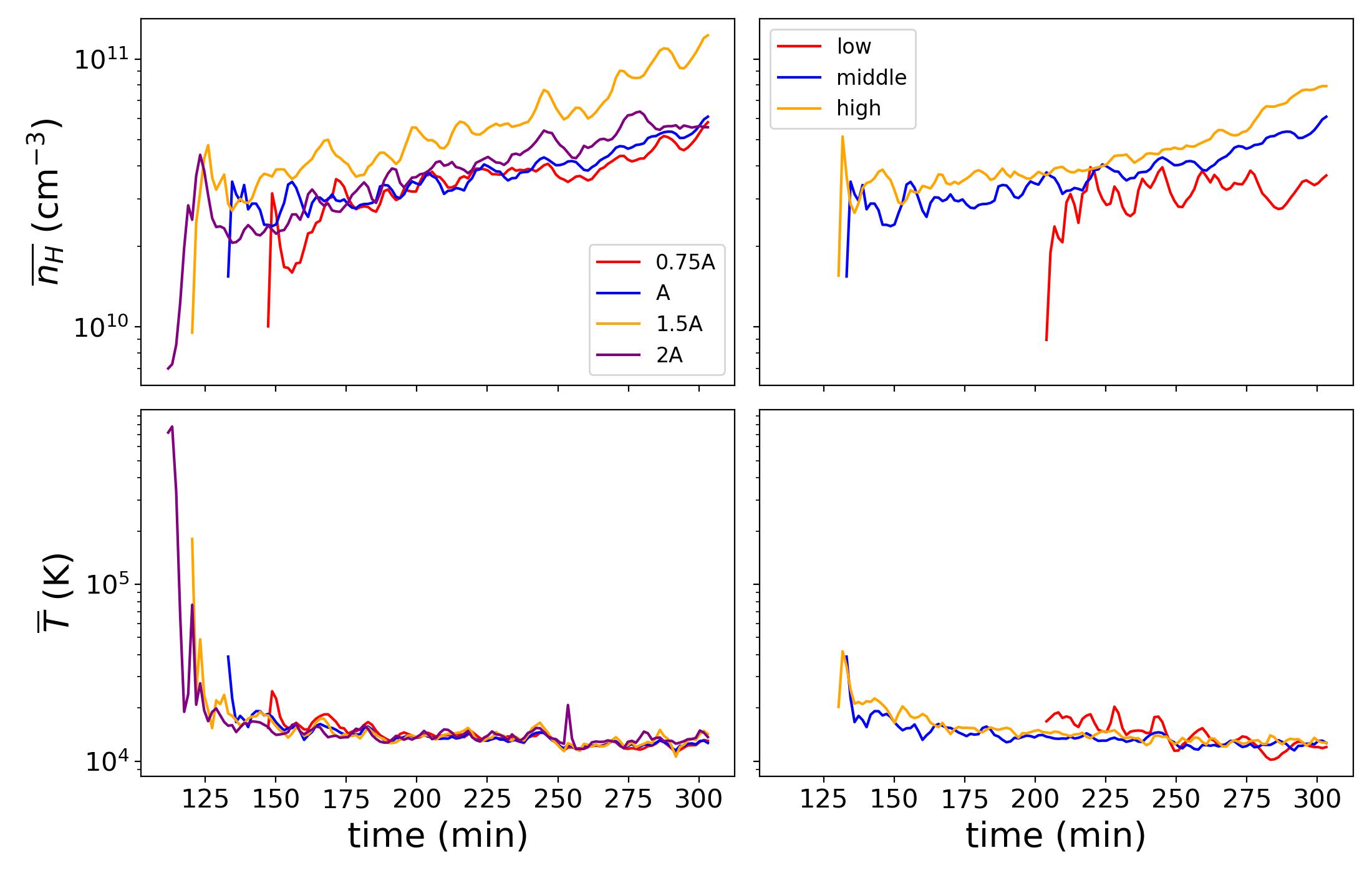}
      \caption{Change in time of the thread-averaged $\overline{n_H}$ (top panels) and $\overline{T}$ (bottom) for all the cases studied here.}
        \label{fig:average_nH_T}
    \end{figure}

    We now detail how the change in the amplitude of the pulses and their height reveals itself in changes of temperature and density. We use the average temperature, $\overline{T}$ and average number density, $\overline{n_H}$ of the threads (averaged throughout only the thread region, identified by the density threshold and the $x$-limits mentioned earlier) from the moment such dense regions start forming. In the top panels of Fig.~\ref{fig:average_nH_T}, the evolution of $\overline{n_H}$ across the different cases is presented. With higher amplitude pulses, more matter is evaporated in the corona. There is an initial sudden jump (in the first 1-3\,min after the first condensation) in $\overline{n_H}$ after which it shows a linear increase, with smaller-scale variations on the shorter timescale. As for temperature changes, after the initial abrupt drop associated with the thermal instability, the average temperature quickly stabilises. After dropping to approximately 10000\,K within the first few minutes, irrespective of the differing amplitude or height cases, only small variations are recorded thereafter. We find minor distinctions between the cases on the basis of the minimum local temperature reached within the threads. Whether due to higher source amplitude or its altitude, we find the minimum thread temperature to lower accordingly. This can be understood through the first order influence of the total thread density as larger magnitudes correspondingly scale the radiative losses by a power of two. 
   
  \begin{table}
      \begin{threeparttable}[b]
    \caption{Condensation rates for each of the cases studied here.}  
    \label{table:cond_rate}      
    \centering                          
    \begin{tabular}{c|c|c}        
    \hline\hline                 
    Case & Condensation & Condensation rate \\  
     & onset time (min) & (g\,cm$^{-1}$\,s$^{-1}$) \\
    \hline\hline                        
      \textit{low} & 202.58 & 2.63 \\   
      0.75A & 145.92 & 4.60\\   
      1A (\textit{middle}) & 131.75 & 5.59\\
      1.5A & 119 & 7.28\\
      2A & 110.5 & 3.72\tablefootmark{a} \\
      \textit{high} & 128.92 & 7.02 \\
    \hline
    \hline
    \end{tabular}
    \tablefoot{
    \tablefoottext{a}{12.24\,g\,cm$^{-1}$\,s$^{-1}$ in the first 36.83\,min.}
        }
      \end{threeparttable}
    \end{table}
    
\subsubsection{Counter-streams and oscillations}    
\label{sec:flows_and_oscillations}

    The time and space varying nature of the footpoint pulses lead to the continuous generation of counter-streaming flows throughout the domain. Every pulse introduces plasma into the corona of slightly different density, temperature and velocity. Such a zoo of motions, as shown in Fig.~\ref{fig:1Avelocities} naturally influences the manner through which the threads form (as in Fig.~\ref{fig:morphology_energies} and \ref{fig:morphology_heights}) for each different case studied here. 
    \begin{figure}
      \centering
      \includegraphics[width=\hsize]{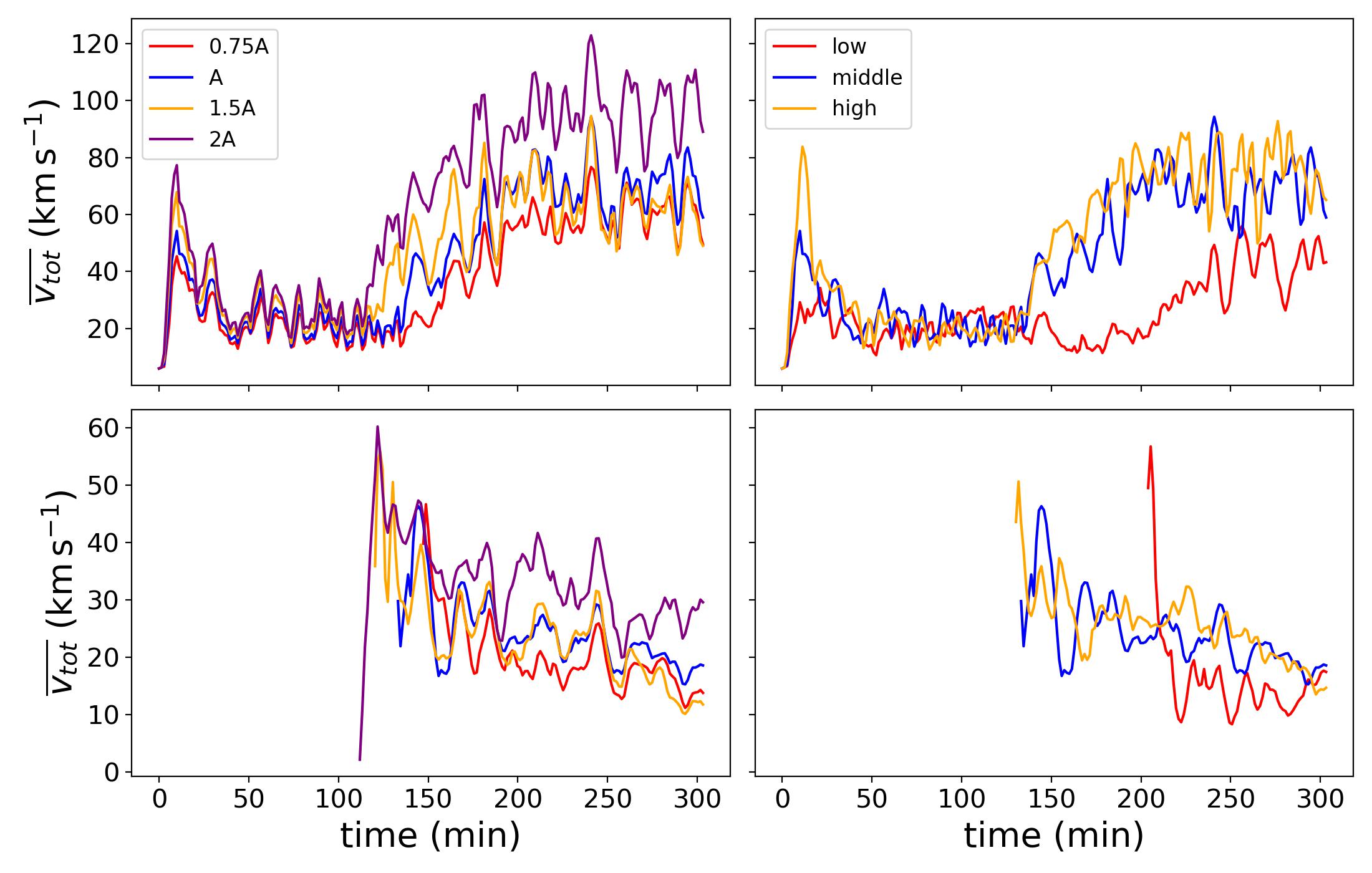}
      \caption{Average total velocity magnitude $v_{tot}$ in the coronal region (top) and in the threaded prominence (bottom).}
        \label{fig:average_vtot}
    \end{figure}
    Figure~\ref{fig:average_vtot} shows the total velocity ($v_{tot}=(v_x^2+v_y^2)^{1/2}$) averaged throughout both the coronal and prominence regions (defined at the beginning of Sec.~\ref{sec:results}). As has already been discussed in previous works \citep{Kucera2003,Arregui2018,Zhou2020}, different temperature flows (observed in EUV and H$\alpha$ images and here referring to the coronal and prominence area) show different ranges of velocities, and in our simulations we notice the same. In addition, we see that a more efficient source, that pushes more matter into the corona (whether due to higher amplitude or its higher position in the atmosphere) shows higher velocities in the coronal, as well as inside the prominence region. For the four cases of different amplitudes, the average total velocities we measure in the corona are between 50 and 120\,km\,s$^{-1}$. For the prominence region, we find the velocities to be in the range between 30 to 50\,km\,s$^{-1}$ initially, followed by a steady decrease. At the final simulated moment, they are around 20\,km\,s$^{-1}$, with the 2A case having a noticeably slower decrease. For the three cases of sources with different heights, the range of the total average velocity inside the coronal region, is between 40\,km\,s$^{-1}$ (corresponding to the \textit{low} case) and up to 80\,km\,s$^{-1}$ (both for \textit{middle} and \textit{high} cases). As for the prominence region, in different height cases, the velocities start in the range between approximately 30 to 45\,km\,s$^{-1}$ and all end again with velocities around 20\,km\,s$^{-1}$. In all of the cases presented here, what is quite noticeable is the increase in the coronal velocities (top panels of Fig.~\ref{fig:average_vtot}) at the moment the condensations begin to form. While the average total velocity within the prominence after that moment decreases with time, the average total velocity in the corona oscillates roughly around this elevated value. What influences this increase and how is it related to the forming condensation we discuss further in Sec.~\ref{sec:Oscillations_counterstreams}.
    
\begin{figure}
      \centering
      \includegraphics[width=0.85\hsize,height=14cm]{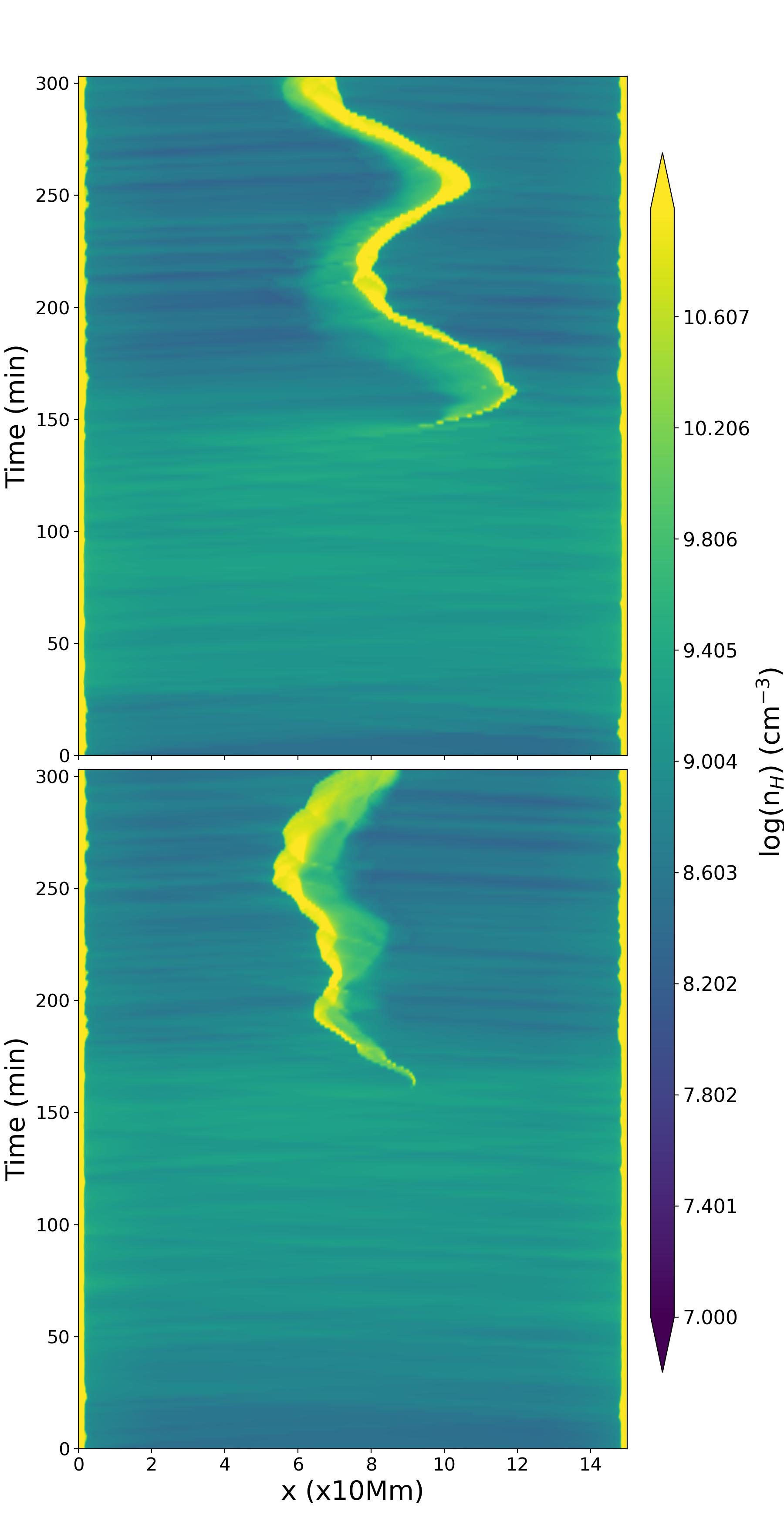}
      \caption{Oscillations of two threads of 0.75A case seen in $x-t$ view at $y=-1.2$\,Mm (top panel) and $y=1.4$\,Mm (bottom panel).}
        \label{fig:0.5A_oscillations}
    \end{figure}

    Another important matter, alongside the average velocities in the coronal and prominence region, are the oscillations of individual threads. In order to show the oscillations, we present in Fig.~\ref{fig:0.5A_oscillations} two cuts along two different threads for the 0.75A case. In the top panel the $x$ cut is at $y=-1.2$\,Mm. This thread shows clearly an oscillation that lasts 55.25\,min between its maximal rightward displacement to $\approx$\,118\,Mm and its minimal leftward position at $\approx$\,77\,Mm, and an approximate velocity of 12.46\,km\,s$^{-1}$. Longitudinal oscillations are usually characterized as either large amplitude oscillations (LAOs) with $v>10$\,km\,s$^{-1}$, or small amplitude oscillations (SAOs) with $v<10$\,km\,s$^{-1}$ \citep{Luna2018}. Accordingly, the oscillations seen here, if we assume they have a half period of 55.25\,min, fall into the category of LAOs with a period similar to the ones reported from observations \citep[][and references therein]{Luna2018,Arregui2018}. During the limited time evolution (150 minutes) we followed, the oscillations of this particular thread show no clear damping. In the bottom panel, a different thread of the same case is shown, namely a cut at $y=1.4$\,Mm. Here, the oscillations are not as clear as with the thread in the top panel, so we refrain from calculating parameters, such as its period, or amplitude. Similarly, in other cases studied here there are examples of threads showing clear oscillations and those showing a more irregular motion. In Fig.~\ref{fig:0.5A_oscillations} we find a change of density noticeable for each thread going from very bright yellow to light green. As mentioned before (Sec.~\ref{sec:reference_case}), since along the whole of its length the thread is not perfectly aligned with the $x$-axis, plotting the parameters along a cut aligned in the $x$-axis, given their finite angle will describe the density transition from inside to outside the thread in a non-symmetric way (cf. Fig.~\ref{fig:xlineplot}).
    
\subsection{Comparison with observations}

\begin{figure*}
      \centering
      \includegraphics[width=\textwidth]{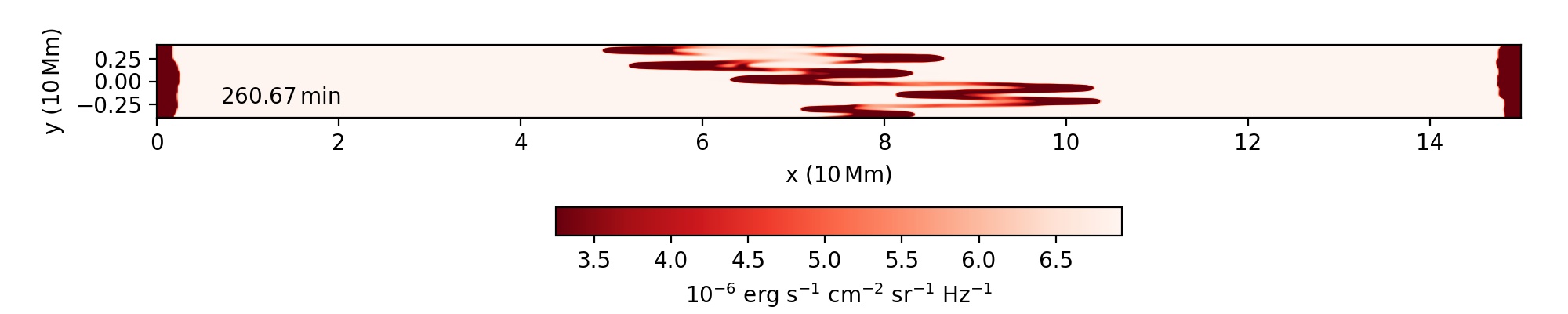}
      \caption{H$\alpha$ representation of the domain of the reference case at $t = 260.67$\,min, with intensity limits as suggested by \cite{Gunar2016}.}
        \label{fig:Halpha}
    \end{figure*}

    Filaments as observed in EUV images typically appear as broad absorption features (particularly in comparison to the Hydrogen H$\alpha$ images), making the analysis and interpretation of individual filament structures difficult. When observing in EUV wavelengths the resulting image is sourced from photoionisation of several elements, plus ionisation within the Lyman continuum, namely H, He and He~{\sc i} \citep{Kucera1998}. Consequently, the appearance of the prominence is a superposition of each multi-thermal column mass which is then integrated along the lines-of-sight, on account of their optical thickness being far less than unity. This then requires the use of extremely high resolutions (numerical or observational) to be discerned \citep{Aulanier&Schmeider2002, Jenkins2022}. As the H$\alpha$ absorption corresponds to a single transition and exhibits an optical thickness closer to unity we explore its advantages in synthesising the threads of our simulation, given their demonstrated prevalence in equivalent observations \cite[e.g.][]{Kuckein2016}. To this end, we follow the implementation of the synthesis method of \cite{Heinzel2015} as reported on in \cite{Claes2020}. The method estimates the H$\alpha$ line absorption coefficient taking into account the local thermal pressure and temperature values. We assume a thread of thickness 1\,Mm. The resulting value is then used in the radiative transfer equation to calculate the H$\alpha$ line intensity presented in Fig.~\ref{fig:Halpha}. The figure shows the domain of the reference case at $t=260.67$\,min. The H$\alpha$ absorption signature forms at lower temperatures than for EUV photoionisation \citep[for comparison see][]{Jercic2022a}, enabling us to clearly see the real fine structure of the prominence threads. If we define the filling factor as the ratio of the prominence area to the total initial coronal area we get a value of $\approx$\,10\% after $t=150$\,min. Similar as in observations, the threads actually take up a small part of the total volume of the coronal plasma \citep[cf.][]{Zhou2020}. From Fig.~\ref{fig:Halpha} additional similar characteristics of prominences are noticed as from observational images. The threaded structure clearly stands out, and threads are elongated across the loop domain with an approximate length of 20-30\,Mm. This compares favorably with the observed thread lengths \citep{Arregui2018}. \\
    The increase in the coronal velocities at the moment the condensations start forming (see Fig.~\ref{fig:average_vtot}) can be interpreted as an indication of the mass exchange between the corona and the condensations, which was also reported in previous observational studies. \cite{OShea2007} studied the variation of intensity (radiant flux) of a number of TR and coronal lines in off-limb loops observed with Coronal Diagnostic Spectrometer onboard SOHO. They noticed a sharp jump in the TR intensity lines coinciding with a sharp decrease in the intensity of coronal lines. Analysing also the velocities associated with those lines they conclude that what they observed was spectroscopic evidence of plasma condensation taking place in coronal loops. The velocities they measured showed a blueshift of up to 100\,km\,s$^{-1}$ in the TR lines (He~{\sc i} and O~{\sc v}). The coronal lines, on the other hand, did not show any significant velocity shift, only the Si~{\sc xii} coronal line showed a velocity redshift of $\approx$20\,km\,s$^{-1}$. They interpret this change of a redshift in the coronal lines to a blueshift in the TR lines as an indication of inflow from the coronal lines to the low temperature TR lines.
    
    Is our stochastic heating enough to create prominences with masses similar to what we measure from observations? From observations \citep{Kucera1998,Williams2013,Carlyle2014} we already know that different types of prominences have different characteristic dimensions (horizontal as well as vertical), that even differ depending on the line of sight of the observation \citep{Mackay2010,Berger2014}. We can estimate the mass of the reference 1A case at the end of the simulation (with a mass per area of $\approx$50000\,g\,cm$^{-1}$). We may assume the height of a thread to be the same as the width ($\approx$1\,Mm) and additionally assume there are more vertically stacked layers comprising the full prominence. In the case of at least 10 layers the total mass will be of the order 10$^{13}$ g, placing our simulation firmly at the lower end of expected masses of prominences.

\section{Discussion}
\label{sec:discussion}
    
\subsection{Global changes in mass and area}

    The formation of the prominence threads is a fast process, happening on a scale of minutes. Once the threads reach their average temperature of $\approx$10000\,K, the optically thin radiative cooling stagnates and there is no further runaway cooling. This average thread temperature drops close to 10000\,K in all cases. On the other hand, there is a difference in the minimum temperature each prominence can reach locally (which is also artificially limited by the numerically imposed minimum of 4170\,K). In general, we find that the stronger the source, and the more mass that is able to evaporate from the chromosphere, the more prominent becomes the thermal instability, enabling threads to reach locally lower minimum values of temperature.
    
    As the mass accumulates faster than the threads grow in area, it leads to a steady increase in their density. The stronger the source (whether due to the amplitude or the height) the faster the rate of mass accumulation, leading to denser prominences, although not necessarily colder on average. If the pulses at the footpoints are strong enough, the mass is eventually lost from the central part of the domain. In real filaments, this process seems to be a major factor in determining the further evolution of the filament. As shown from observations \citep{Bi2014, Jenkins2018}, modeling \citep{Jenkins2019} and numerical works \citep{Fan2020} the decrease in mass actually means a decrease in the force restraining the filament, which leads to its raise and further expansion, often also eventually to eruption. Our 2A case experiences such drainage of plasma from the top of the arcade back to the chromosphere. It nicely depicts the circulation of plasma of the threads. We see evaporation from the chromosphere (cold plasma heated up), condensation in the corona (heated plasma cooling down) and eventually its drainage back again into the chromosphere (in the form of coronal rain). Further on, the mass balance affects the density and with it the motion of the threads, as we see from Fig.~\ref{fig:average_vtot}, the increase in mass/density influences the decrease of the velocity during the evolution. We can speculate that it is the particular combination of the plasma evaporation in the chromosphere and the magnetic topology that determines the balance between condensation and drainage, resulting in more or less permanent structures \citep[prominences or coronal rain,][]{Liu2012}. 
    
    We can refer here to a previous 1D study done by \cite{Xia2011} and make a comparison. These authors studied how the change in amplitude of a steady, localized heating affects the growth rate and onset time of the prominence. As in our study, the onset time decreases as the amplitude increases, which is expected. 
    Quite similar to their length growth, the area here shows a more intense growth at the beginning (first 10-20\,min) after which it reaches a more steady increase. 
    Unlike their 1D results, where there was no drainage and the filament kept continuously increasing in length, here we see the growth also depends whether drainage is present or not, and on the relation of that drainage with the condensation. 
    %
    On the other hand, \cite{Xia2011} showed that the change of the growth rate with the amplitude of the source is not a simple linear relation, and instead reached a maximum value since stronger heating eventually means more difficult cooling. Also, stronger heating implies more evaporated plasma which leads to higher compression and again lower growth rate. 
    Of course, a direct comparison is not straightforward as the 1D case had a steady heating amplitude, while in our 2D case we have multiple pulses. 
    Furthermore, their steady background heating is localized over 10\,Mm, while in our case we have the $x$ and $y$ length scale of only 1.5\,Mm. Additional factors that likely play a role here are the differences in the adopted arcade topology, where e.g. the loop length can influence the time the condensation happens \citep{Cargill&Bradshaw2013, Kaneko2017}. The loop implemented by \cite{Xia2011} extended some 260\,Mm with a significantly smaller dip of the middle section (only 0.5\,Mm). 

\subsection{What role plays the height of the impulsive heating?}

    Comparing the cases of different heights, we point here to the following: the lower the pulse is in the atmosphere, the more energy it needs to be comparable to a pulse positioned higher up. Between the \textit{high} and the \textit{middle} cases there is no significant difference in the resulting condensations (especially looking at the total area and the total mass values). The \textit{low} case, on the other hand, shows significantly slower accumulation of mass and area. 
    As the atmosphere is gravitationally stratified, pulses at lower heights do work against the mass piling on top that is exponentially varying. Pulses "buried" more deeply in the chromosphere have a much harder time lifting the material into the corona compared to the pulses already within the corona. This explains why there is a significant difference between the \textit{low} and the \textit{middle} and a not so significant difference between the \textit{middle} and the \textit{high} case, even though the actual increase in height is the same. 
    
    By varying the height of the source, one could in principle also achieve prominences by direct injection, as in the 1D study by \cite{Huang2021}. However, we see that the source height is not the only major factor playing a role here, and our 2D study did not find evidence for injected prominences in the parameter regime we studied.
    
\subsection{Counter-streams and oscillations}
\label{sec:Oscillations_counterstreams}

    From the results of our simulations we can conclude that the counter-streams and their properties are dictated by the heating sources lower down in the Sun’s atmosphere. 
    In general, we notice that the threads that originate from weaker/lower pulses develop slower in comparison to the threads originating from stronger/higher positioned pulses (Fig.~\ref{fig:morphology_energies} and~\ref{fig:morphology_heights}). Further on, the pulses also affect the velocities we measure in our domain, in particular the average velocities in the coronal region (Fig.~\ref{fig:average_vtot}). The difference is clearly seen when we change the pulse amplitude. When changing the height of the pulses the difference in the average total velocity in the corona between the \textit{high} and the \textit{middle} is not strongly pronounced. However, in the \textit{low} case, even though we decreased the height by the same amount as between the \textit{high} and the \textit{middle} the resulting velocities are significantly lower. We can also notice how in all of the cases the average total velocity in the threads continuously decreases with time, although the 2A case is an exception here. When the drainage starts (at around 150\,min the prominence has a sharp drop in mass) the decrease of the average $v_{tot}$ of the prominence slows down. Relating the two, the continuous decrease in the average velocity of the prominence may well be caused by the increase in the average density (i.e. mass) of the threads, as noticed in \cite{Luna&Karpen2012}. They report on the damping of oscillations of a prominence as a result of ongoing accretion of mass by the condensation. In a similar manner we can explain the jump that the coronal velocities (top panels of Fig.~\ref{fig:average_vtot}) experience upon the onset of condensation. When the localized heating starts ($t=0$) the plasma from the footpoints is evaporated into the coronal part of the arcade and it increases its overall mass. At a particular moment the condensations start forming, the mass of the coronal part is depleted and followed by an increase of the average velocity of the flows of the coronal plasma. 
    
    As for the thread oscillations, they show differences in different cases, and even for different threads of the same case. 
    The oscillatory motions we observe are a consequence of an interplay of multiple factors: gravity playing the role of the restoring force \citep{Luna&Karpen2012,Luna2022}, the contribution of the pressure gradient force along a rigid magnetic field topology \citep[similar as in][and also seen in Fig.~\ref{fig:xlineplot}]{Zhang2019}, and the thermodynamic changes induced by the pulses. 
    While some threads do not show a clear oscillatory pattern (bottom panel of Fig.~\ref{fig:0.5A_oscillations}), others show very clear oscillations (top panel of Fig.~\ref{fig:0.5A_oscillations}). It seems that for some threads the flows support and contribute positively to the oscillations, while for others the flows change their initial trajectory and disturb the threads from their expected oscillatory behaviour. In a recent paper by \cite{Ni2022} the authors analyzed decayless oscillations of a filament observed near an active region on July 5th, 2014. They conjecture that the decayless oscillations are caused by quasi-periodic jets happening as a result of intermittent reconnection at the footpoints of the flux rope supporting the filament. They noticed that the period of a filament driven in such way differs from the pendulum model and that it directly depends on the period of the jets. 
    This jet can either act in the same direction as the restoring force (decreasing the period) or in the opposite direction of the restoring force (prolonging the period). 
    In their case there was a single thread along a single magnetic field line (1D simulation) and multiple equally time-spaced jets coming from one footpoint. Here we have multiple field lines with a finite number of threads along them. The driving is constantly present but with a random interpulse interval in the range between 100 to 300\,s coming asymmetrically from both sides of the arcade. In the example on the top panel of Fig.~\ref{fig:0.5A_oscillations} we do observe, similar to \cite{Ni2022}, oscillations without decay, but also find clearly disrupted oscillatory behaviour. There is hence no clear dependence of observed periods with the interpulse interval of the pulses that we can infer. 
    
    Lastly, in the 2A case we noticed short-lived (temporary) small-scale structures resembling billows of Kelvin-Helmholtz instability (KHI), that result from shear flows with an additional difference in densities between them. However, due to the strong magnetic field in our domain this KHI is suppressed from fully developing.

\section{Summary and Conclusions}
\label{sec:conclusion}
    In this work we investigated the influence of randomized and localized (in time and space) heating on the fine-structured condensations (prominences with threads) in a given magnetic arcade setup. In a previous, purely adiabatic study \citep{Jercic2022} of threaded prominences (where we inserted the prominences, rather than formed them), we showed that the influence of a single energetic pulse is considerable, especially when studying oscillations of multi-threaded solar prominences. Here we show that superposed, relatively small pulses (a result of many small-scale reconnection events or nanoflares), can cause naturally fine-structured plasma condensations. These perturbations have a great influence on the dynamics of individual threads, on their thermodynamics and even possibly the extent of their existence. We list more important conclusions:
    \begin{enumerate}
        \item The average temperature of the threads that form does not show strong differences between cases that vary in either amplitude or height of the random pulses. On the other hand, the actual minimum of temperature does  decrease with stronger/higher source as a result of more mass evaporated from the chromosphere, creating denser condensations and hence a stronger effect of the radiative cooling (more efficient thermal instability).
        \item Considering that the sources are constantly present throughout our simulation, they drive a continuous and steady increase in mass and density. We quantified the condensation rates of our multi-threaded prominences (Table~\ref{table:cond_rate}) and assuming threads of height similar to their width ($\sim$1\,Mm) with at least 10 layers, we get condensation rates in the range of (2.63-12.2)$\times$10$^9$\,g\,s$^{-1}$. Those values are comparable to the condensation rates measured from observations by \cite{Liu2012} in the first 3\,hr of prominence formation.
        \item The height and the amplitude of the pulses at the footpoints affect the topology of the prominence. Higher/stronger pulses create threads faster and they become longer in comparison with the lower positioned/weaker pulses.
        \item We showed a linear increase of the condensation rate with the amplitude of the source. As we only varied amplitudes by modest factors, more comprehensive studies are still needed in order to properly understand the effects of stochastic heating in multi-dimensional settings.
        \item As the arcade atmosphere is gravitationally stratified, the mass changes exponentially with height along both footpoints, making the perturbation within the corona weaker the lower this perturbation is in the atmosphere. As a result the effect of the source drastically decreases with height.
        \item The pulses at the footpoints are the source of observed counter-streams. Together with the threaded prominence, they result in a banded flow in the corona. The counter-streams strongly affect the evolution of the threads. The flows are dictating the average velocities and general motion of the threads. The streams themselves are influenced by the amplitude of the pulses and their height.
        \item The motion of the threads is the combined result of the gravity force, thermal pressure gradient and randomly distributed pulses at the footpoints. As a result, some threads show clear oscillations while others show only irregular, erratic or one-sided motion.
    \end{enumerate}
    Our work shows that variations in the heating source can influence observed prominence threads dramatically, especially the motion of the threads. The findings on prominence oscillations previously done in 1D simulations are shown to be more complex in actual 2D counterparts. Our study focused on amplitudes and heights of random pulses, other parameters still need to be surveyed: the length scales in our $x$ and $y$ directions, the duration and time between the pulses. To study prominence threads and their formation, high resolution is crucial, especially for characterizing the width of the formed condensations. The new telescope EUVST (EUV High-throughout Spectroscopic Telescope) onboard the SOLAR-C mission will allow seamless observations of the Sun in a temperature range starting from chromospheric to coronal temperatures with unprecedented resolutions (0.4 arcsec at maximum). This spectroscopy will allow further studies on the energy exchange in the Sun's atmosphere and a more successful comparison of observational data and simulation results. Moreover, as new data about the Sun's magnetic field will be arriving with new space and ground-based instruments (Solar Orbiter, Daniel K. Inouye Solar Telescope), we will further contrast our simulation models with observations. To that end, we provided proxy H$\alpha$ views. More realistic radiative transfer treatments are left for future work. 
\begin{acknowledgements}
      We thank the referee, Peter Cargill for his comments that helped improve the paper. VJ acknowledges funding from Internal Funds KU Leuven and Research Foundation – Flanders FWO under project number 1161322N. This project received funding from the European Research Council (ERC) under the European Union’s Horizon 2020 research and innovation program (grant agreement No. 833251 PROMINENT ERC-ADG 2018), and is supported by Internal funds KU Leuven, project C14/19/089 TRACESpace and FWO project G0B4521N. Visualisations used \href{https://www.paraview.org/}{ParaView}, \href{https://www.python.org/}{Python} and \href{https://yt-project.org/}{yt}. The resources and services used in this work were provided by VSC (Flemish Supercomputer Center), funded by the Research Foundation - Flanders (FWO) and the Flemish Government. We thank Y. Zhou, J. Hermans, N. Claes and J. Jenkins.
\end{acknowledgements}

%
%

\bibliography{references}{}
\bibliographystyle{aa}
\end{document}